\documentclass[jkps,fleqn,showpacs,showkeys]{revtex4}
\usepackage[pdftex]{graphicx}
\usepackage{amssymb}
\usepackage{extpfeil}
\usepackage{amsmath}
\usepackage{bm}

\begin{document}
\setcounter{page}{1}
\title[]{ Full simulation on the dynamics of auditory synaptic fusion: Strong clustering of calcium channel might be the origin of the coherent release in the auditory hair cells   }
\author{Jaeyun \surname{Yoo}}
\author{Kang-Hun \surname{Ahn}}\email{ahnkanghun@gmail.com}
\affiliation{Department of Physics, Chungnam National University,
 Daejeon, 305-764, Republic of Korea, \\
 Hearing Loss Research Laboratory, Deep Hearing Corp.}
\date[]{Received !Month !day 2025}

\begin{abstract}
The precise timing of synaptic transmission in auditory hair cells is important to hearing and speech recognition. Neurotransmitter release is an underlying step in translating sound. Thus, understanding nature of the synaptic fusion is key to understand the hearing mechanism. Extraordinary large excitatory postsynaptic currents (EPSCs) have been observed in the auditory hair cell synapse, and its origin has been controversial. It is not known yet whether the size and shape of the EPSCs are  results of a big vesicle or many small vesicles.   We report our numerical
simulation of the vesicular fusion process from calcium channel process to
the generation of EPSC currents. Our numerical experiments indicate that
the origin of the large EPSC with its mysterious form is close to the scenario of the multivesicular release.  The large EPSCs might be triggered by strong calcium channeling of the calcium channel clusters.
\end{abstract}



\maketitle

\section{INTRODUCTION}

\begin{figure} [hbt!]
\centering
\includegraphics[width=0.9\textwidth]{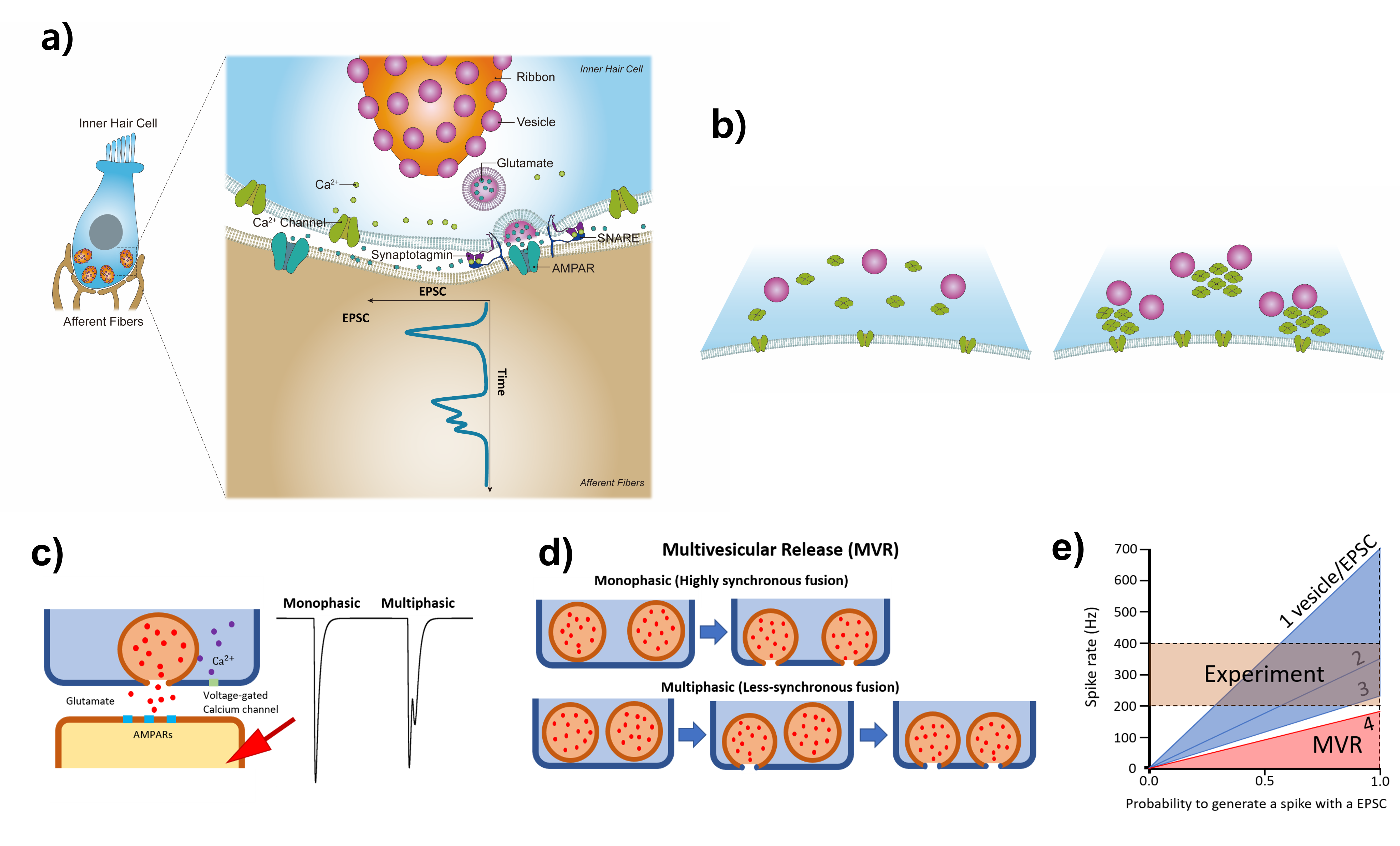}
\caption{a) Schematic of the structure of auditory neurons and inner hair cells. b) Schematic of calcium channel clusters. Calcium channels have been often assumed to be uniformly distributed within the cluster area (left), but it might be also possible that a few channels gathered locally even in the cluster area (right). c) Calcium channels activates depending on the voltage of the inner hair cell. A vesicle is fused when it attaches to calcium ions. From the fused vesicle, glutamate diffuses through synaptic cleft and activates AMPA receptors, which generate monophasic and multiphasic EPSCs. d) MVR model of interpreting monophasic and multiphasic EPSCs. e) The maximal sustained synaptic vesicle release rate per active zone is approximately 700 Hz and on average six vesicles compose an EPSC in the MVR scenario. Because one EPSC can trigger only one spike, a maximum spike rate of $\sim$120 Hz is predicted. Yet, each IHC synapse can drive spiking at 200-400 Hz during ongoing stimulation in vivo, where IHC depolarization likely is weaker.
}
\label{figure1}
\end{figure}

Mammals' auditory hair cells convert external sounds into electrical signals in auditory neurons (Fig. \ref{figure1}a ). This neural signal is quickly transmitted along the nerve fiber to the central neural system with precise timing, which is very important to human perception of speech.
Intensive research on the machinery of the neural synaptic transmission \cite{Redman, Gulyas, Lawrence, Silver, Murphy, Biro, Tong} has shown that there is a mass of protein called {\it ribbon} and hundreds of vesicles hang around it \cite{Nouvian}.
At each presynaptic release site, vesicles carrying a neurotransmitter move onto the readily-release pool through the ribbon and are probably released simultaneously in a coherent manner (MVR: Multi Vesicular Release). Otherwise, probably multiple vesicles form a large vesicle before the release (UVR: Uni Vesicular Release). 
The UVR scenario seems to be evidenced 
by the fact that the experiments on the auditory synaptic current show that a large portion of the EPSC show large charge and amplitude \cite{Glowatzki1, Grant, Nikolai}.
However, UVR, as the origin of the strong EPSCs, is challenged because it cannot explain why the strong EPSCs appear in two different forms -  a single peak (monophasic EPSC) and a signal with multiple curves (multiphasic EPSC) \cite{Glowatzki1, Grant}. 

Recently, there have been also intensive studies on the interaction between calcium channels\cite{Wong,Liu}. 
The calcium channels might move due to the inter-channel interaction such as van der Waals force and/or entropic force \cite{Asakura1, Asakura2, Moly}. 
In this work, we find that the positioning of the calcium channels affects the vesicle fusion processes. 
Here, we propose that the two mysterious features, strong EPSCs and calcium channel clustering are fundamentally intertwined and can be understood in a unified manner.  Here, we demonstrate through our numerical simulation that the locally clustered calcium channels (Fig.\ref{figure1}b ) may simultaneously release calcium ions, resulting strong EPSCs.

It has not been clear why some EPSC peak heights are much
larger than small EPSC peaks and why they are in two different forms (Fig. \ref{figure1} c ).
Both the two scenarios UVR\cite{Nikolai} and MVR\cite{Grant} cannot fully explain the origin of the large EPSC phenomenon.
In the UVR scenario, multiphasic signals might be attributed to {\it flickering} pores when vesicles release neurotransmitters. However, it is difficult to imagine any physical mechanism for the incredibly frequent vibration of the vesicle pore.
If we try to explain the observed sizes of EPSCs within the UVR scenario, the size of the vesicles should be larger than the observed sizes of the vesicles \cite{Grabner, Nikolai, Neef2}.
The MVR scenario might be suitable to explain the large EPSCs if there is a mechanism
that leads to coherent vesicle fusions simultaneously. 
In this paper, we report our numerical simulations on the vesicle fusions considering various
calcium channel locations, calcium channel gating, calcium ion diffusion, sensor-ion interaction, and
glutamate diffusion and its receptor.

\begin{figure} [hbt!]
\centering
\includegraphics[width=0.7\textwidth]{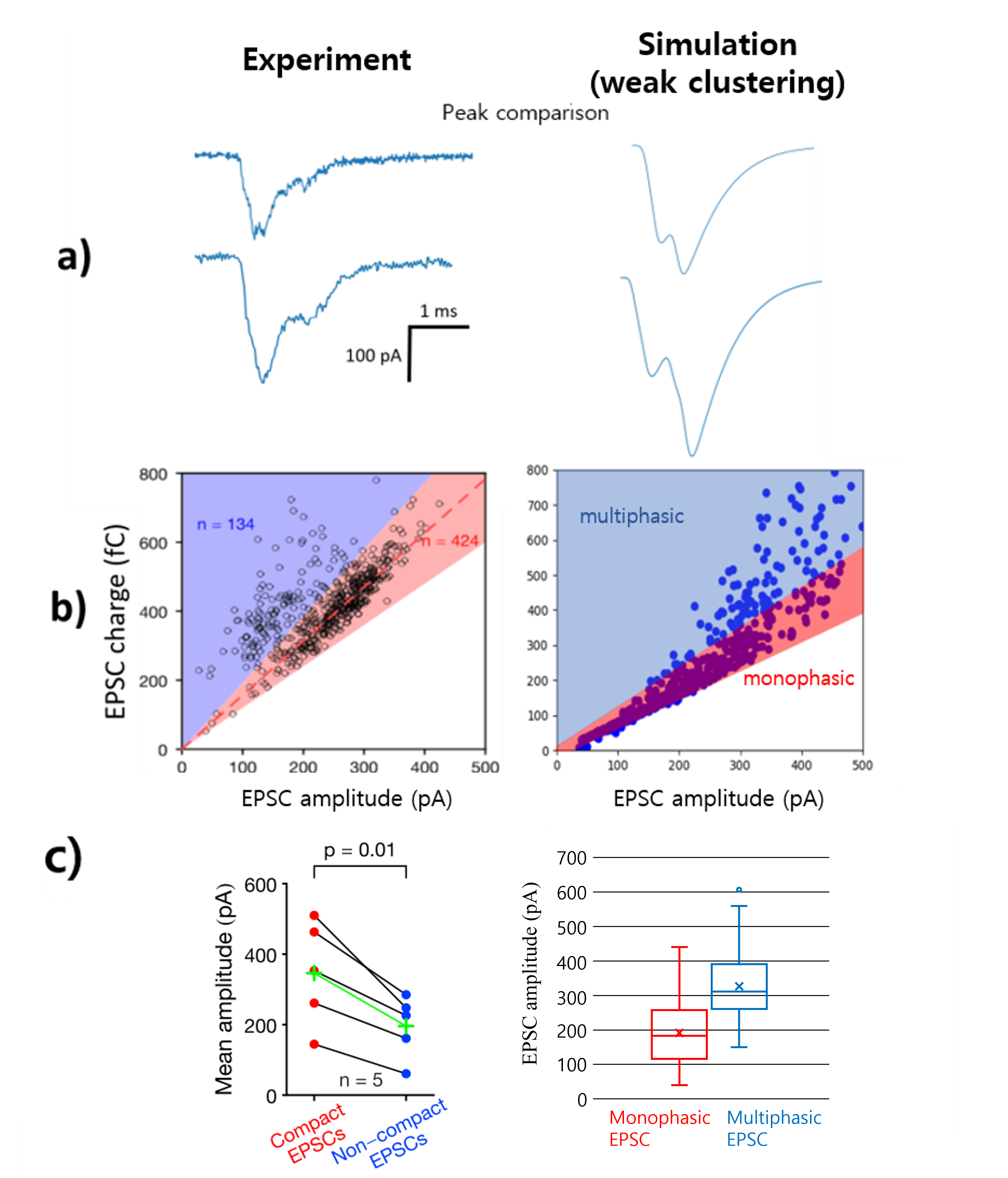}
\caption{ Feature comparison between experimental results and simulation. a) In experiments, the first peak is the largest but, simulation results show some randomness. b) Experiments show that the amplitude of multiphasic EPSCs is smaller than monophasic EPSC. However, in the simulation, EPSCs with smaller amplitude ($<$3 vesicles) have monophasic, and larger amplitude ($>$5 vesicles) have multiple peaks.
c) (Left) Experimental data reproduced from Ref\cite{Nikolai}. Comparison of the average amplitude and charge between experiment and calculation data.
}
\label{figure2}
\end{figure}

\section{Results and Discussion}

In the MVR scenario, two different shapes of EPSCs (monophasic and multiphasic) seem to be probably understandable by introducing the degree of synchrony of the vesicle fusion.
 A possible mechanism is that the shapes of neural signals differ due to the interval of timing of the emitted vesicles. In other words, monophasic EPSCs occur when multiple vesicles are released at the same time, meanwhile multiphasic EPSCs are generated with less synchronization (Fig. \ref{figure1} d) \cite{Grant}. 
 However, the observed spike rates are seemingly contradictory
within the MVR scenario. (See Fig. \ref{figure1}e )
 The number of vesicles for the stimulation state with a depolarization level of -14 mV is about 700 per second from capacitance measurement\cite{Pangrsic}. 
 A typical strong EPSCs are about six times larger than the smallest EPSC, which implies that about six vesicles are fused simultaneously if we are within the MVR scenario \cite{Grant}.
  However, the observed spike rates measured from the afferent synapses range from 200 to 400 spikes/s (Fig.\ref{figure1}  e) ) \cite{Taberner}, which is much higher than 700 / 6 $\approx$ 117.
This consideration leads us to find out whether strong EPSCs can be generated by a small number of vesicles fewer than six vesicles. 

In Fig. \ref{figure2}, we report our computational experiment that simulates vesicle fusion, calcium and neurotransmitter diffusion, and EPSCs.
The relationship between EPSC amplitude and the EPSC charge, obtained by integrating the EPSC over time, can be used to infer whether the EPSC is monophasic or multiphasic. 
As a characteristic of the neural signal, monophasic EPSC has a relatively large average amplitude and a wide range of amplitude of about 50 to 600 pA. Multiphasic EPSC is distributed below the average of monophasic neural signals and has a relatively large charge \cite{Nikolai}.
For a given EPSC charge, a larger EPSC peak amplitude suggests a more monophasic profile. Fig.\ref{figure2}b (left panel) presents the experimental data, while Fig.\ref{figure2}b (right panel) shows the corresponding simulation results. 
The experimental and simulated data exhibit qualitatively similar patterns.
However, upon closer examination, a discrepancy between the experimental results and the simulation can be observed. As shown in Fig. \ref{figure2}c, the amplitude of the multiphasic responses (non-compact EPSCs) are lower than those of the monophasic responses (compact EPSCs) in the experimental data, whereas our theoretical prediction (in the right panel) suggests the opposite.

\begin{figure} [hbt!]
\centering
\includegraphics[width=0.7\textwidth]{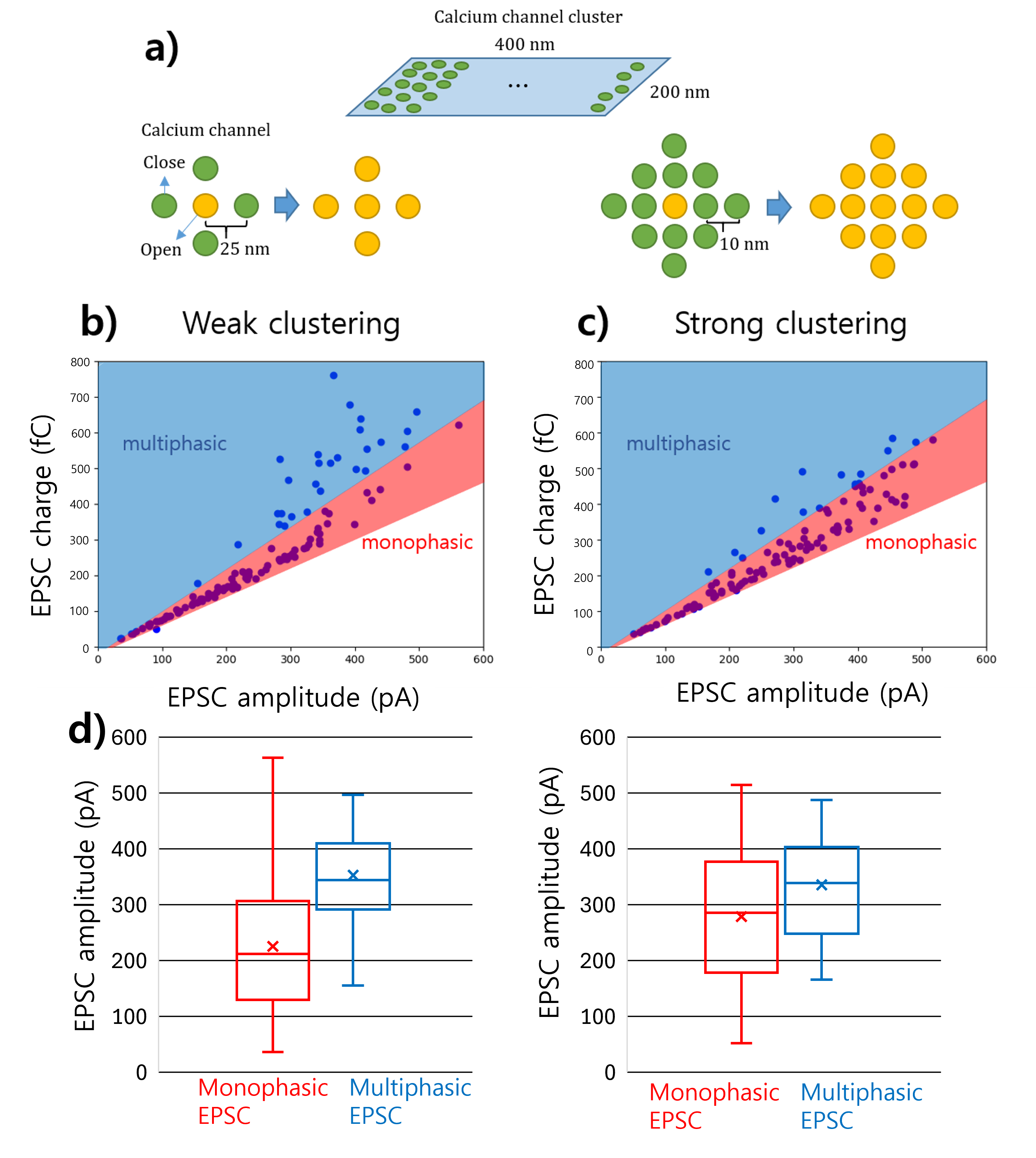}
\caption{
a) The structure of calcium channel clusters. Both conditions when calcium channels are sparse and dense are calculated. Also, when one channel is opened, the neighboring channels also open.
b), and c) The distribution of amplitude and charge of EPSCs for sparse and dense channel cluster condition.
d) Comparison of the average amplitude and the charge of the EPSC for weak calcium channel clustering (left) and
strong calcium channel clustering.}
\label{figure3}
\end{figure}

Recent studies have been conducted on the interaction of ion channels, suggesting the possibility of local cohesion due to the attraction between channels due to the entropy force and the protein movement \cite{Moly}.
We therefore conduct simulations under the hypothesis that attractive interactions among calcium channels could give rise to strong clustering.  Fig. \ref{figure3}a illustrates the spatial distributions of the calcium channels used in our simulations. The left panel shows a configuration with weak clustering, and the right panel depicts a distribution exhibiting strong clustering. The simulation results shown previously in Fig. \ref{figure2} were obtained under the assumption of weak clustering. Our initial choice of a 25 nm inter-channel distance, corresponding to weak clustering, was based on the average calcium channel density observed in experimental studies.
However, if calcium channels interact in such a way that they aggregate more tightly, this could lead to localized regions of high calcium concentration, which in turn might significantly alter the pattern of vesicle fusion. Indeed, as shown in the comparison between Fig. \ref{figure3} b, c, and d our simulations demonstrate that strong clustering leads to increased amplitudes of monophasic EPSCs.

Although the amplitude of monophasic EPSCs does not yet exceed that of multiphasic EPSCs in the current simulation results, it is evident that increased clustering of calcium channels leads to a clear enhancement in the amplitude of monophasic EPSCs.
If more strong clustering of calcium channels is possible, our simulation results suggest that multivesicular release may not only generate large EPSCs but also give rise to high amplitudes of monophasic EPSCs.


\section{Method}

The mechanical stimulus applied to the stereocillia of auditory hair cells
induce the K$^{+}$ influx through mechanoelectric transport channels.
The increase of the membrane voltage due to the pottasium influx
is the input signal in our numerical experiments.
We model presynaptic region as 80 voltage-induced calcium channels which
are in a 4 by 20 array and about 20 vesicles are on the calcium
channel cluster ( Fig.\ref{figure3}a ). 
This parameters come from the experiments on the postnatal rats\cite{Nouvian, Magistretti}.

 In our simulation, there are four main processes for EPSC generation:\\ 
 i). calcium channel gating, \\
 ii). calcium diffusion, \\
 iii). sensor-ion interaction,\\ 
 iv). AMPA receptor model.
 
 From below, we will describe the detail of our model for each process. The controversy on the two types of the EPSC ( monophasic and multiphasic  ) is about
the experiments on rats in spontaneous conditions, so we take all parameters in our simulations from
the experiments in the spontaneous condition without stimulus.

\subsection{Calcium channel gating}

In an inner hair cell (IHCs), there are the 5-30 ellipsoidal ribbons\cite{Moser}.
Under the ribbon structure, there are vesicles in  two-strand structure   
 and calcium channels in the form of stride-type cluster\cite{Vincent, Wong}.
 Voltage-gated L-type calcium channels play an important role in the IHC. Especially, Cav1.3 channels, which are a kind of L-type channel, are exclusively located in mammals. The diameter of calcium channels is around 15 nm, and that of vesicles is around 40 nm.
 Calcium channels have three states: rest, activate, and open. The gating dynamics of the calcium channels can be treated by a three-states Markov chain model such as,

\begin{eqnarray}
C_1 \xrightleftharpoons[k_-]{2k_+} C_2 \xrightleftharpoons[2k_-]{k_+} O.
\label{calcium_rate}
\end{eqnarray}

 The calcium current from an IHC is fitted with the following empirical equation with the maximum calcium current $I_{\rm max}$ and the
activation time constant $\tau_{\rm act}$\cite{Scharinger,Lyshevski},

\begin{eqnarray}
I(t) = I_{\rm max}(1-\text{exp}(-t/\tau_{\rm act}))^2,
\label{calcium_current_fit}
\end{eqnarray}

\begin{eqnarray}
I_{max} = \frac{g_{max}(V_m-V_{rev})}{1-\exp(\frac{V_{\frac{1}{2}}-V_m}{S})},
\label{calcium_current_max}
\end{eqnarray}

where the form of this equation is the solution of the three-state rate equation corresponding
to Eq.(\ref{calcium_rate}).
The activation time constant $\tau_{\rm act}(V_{m})$ have been measured within the
whole-cell patch clamp method mostly for constant membrane voltage $V_{m}$.

While the value of activation time near the resting membrane voltage is not known, as its amplitude is too small, we get the value for
the resting condition based on empirical formula which was found 
 in Ref.\cite{Johnson},
\begin{eqnarray}
\tau_{\rm act}(V_{m})= \frac{4}{1+\exp(\frac{V_{m}+0.054 {\rm mV}}{0.009 {\rm mV}})}+0.34 {\rm ms}
\label{act_time_fit}
\end{eqnarray}

The saturated open probability of the calcium channels is given by the following formula:
\begin{eqnarray}
P_{o}(\infty) = \frac{I_{max}}{i_{ch}\times N_{ch}}=\frac{I_{max}}{g_{ch}\times V_{m} \times N_{ch}},
\label{channel_open_prob}
\end{eqnarray}
where $i_{ch}$ is the calcium current for single channel, and $N_{ch}$ is the total number of channels in an IHC. 
The single channel calcium current $i_{ch}$ follows the Ohm's law with the conductance of $g_{ch}=14.7 pS$ \cite{Zampini}.

The rate constants ${k_+}, {k_-}$ in Eq.(\ref{calcium_rate}) can be obtained from their relation to the activation time constant $\tau_{\rm act}$ and the saturated open probability $P_{o}(t=\infty)$ (\cite{Wong}),
\begin{eqnarray}
\tau_{\rm act} &=& \frac{1}{k_{+}+k_{-}} \\
P_{o} (\infty) &=&\left(\frac{k_{+}}{k_{+}+k{-}}\right)^2  .
\end{eqnarray}

We simulate the calcium channel opening using the three-state model in Eq.(\ref{calcium_rate})
with the parameters aforementioned in this section.

\begin{figure} [hbt!]
\centering
\includegraphics[width=1.0\textwidth]{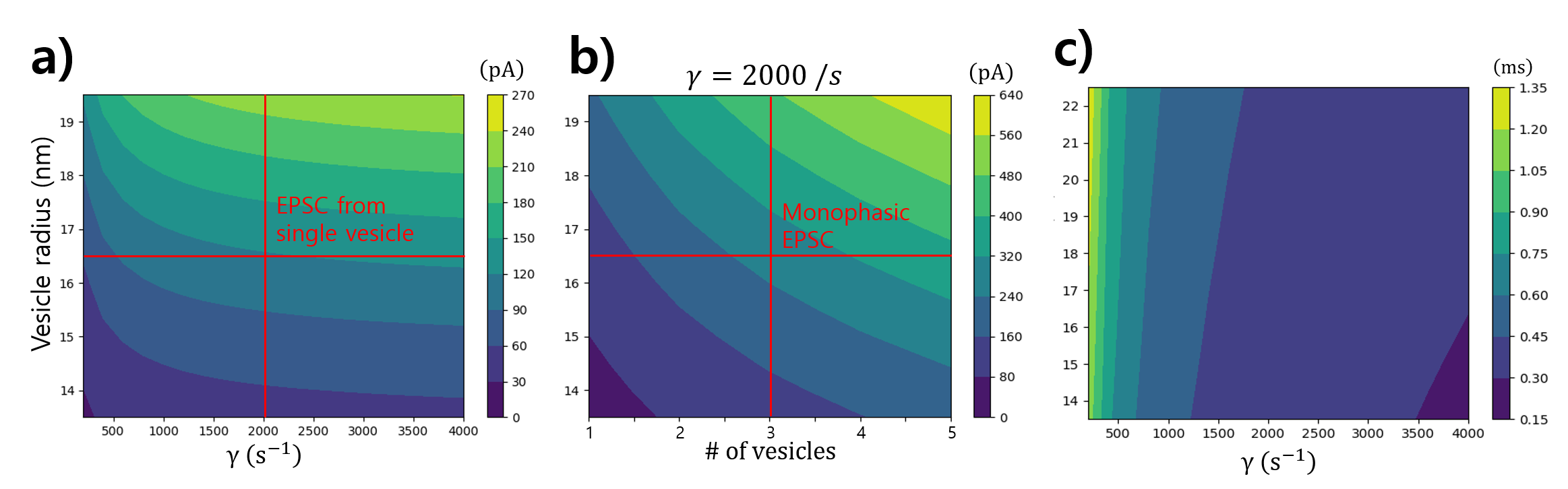}
\caption{
a) The plot of EPSC amplitude when the pore opening rate and vesicle radius change. Over $\gamma = 2000 s^{-1}$, EPSC amplitude barely increases. For a average vesicle size (16.5 nm),  EPSC amplitude is 150 pA.  
b) The plot of EPSC amplitude when the number of vesicles and vesicle radius change. Fusion of three vesicles generates 250 pA, which is 
the average EPSC amplitude.
c) The plot of the interval time between vesicle fusion and maximum EPSC amplitude. The delay time barely decreases from over $\gamma = 2000 s^{-1}$.
}
\label{figure4}
\end{figure}

\subsection{Calcium ion diffusion}

When the calcium channel opens, the calcium ions diffuse quickly in the presynapse which affects vesicular fusion.
To describe the vescular fusion timing accurately, we need to describe the calcium ion diffusion in more precisely.
In the presynaptic region, calcium ions are not free to diffuse. Most calcium ions are bound by protein buffers and the interaction of calcium ions, and buffers can be considered by the following relationship.

\begin{eqnarray}
Ca^{2+} + B_i \xrightleftharpoons [k_i]{k_{-i}}  CaB_i, \\ \nonumber
~~~~~~~~~~~~~~~~~~~~~~~~~~~~~~~~~~~~~~~i=1,2,...,N
\label{calcium_bind}
\end{eqnarray}

where $B_i$ is the i-th specie of the buffers and N is the number of the species. 
The equation for the diffusion of calcium ions in the presence of buffers is described with a diffusion term and linearized buffer interaction \cite{Naraghi}. The calcium concentration becomes steady-state in submilliseconds, so it has been convenient to use the steady-state calcium concentration\cite{Naraghi, Wong}. However, since it is important to simulate accurate vesicle fusion timing in this work, we need to describe transient behaviour of time-varying calcium concentration. From Ref. \cite{Naraghi}, we consider a transient solution by applying the additional term to represent an environment similar to the presynaptic zone with a ribbon structure.
The Calcium concentration levels were estimated by treating open presynaptic calcium channels as calcium point sources and by using a linearized buffer approximation:

\begin{eqnarray}
\frac{\partial y}{\partial t} = A\cdot y + D\cdot (\frac{\partial ^2}{\partial r^2}+\frac{2}{r}\frac{\partial}{\partial r})y,
\label{diffusion}
\end{eqnarray}
where $y$ is a vector composed of each $CaB_i$ and calcium concentration, which is denoted by $y=([CaB_1],...,[CaB_{N}],[Ca^{2+}])^T$. $D$ is the diagonal matrix of the diffusion coefficients. Following Ref \cite{Wong}, we consider the endogenous calcium buffers: calretinin (CR), calbindin (CB), parvalbumin (PV), and ATP. Calcium and calcium buffer concentration, their binding and unbinding rates, and diffusion coefficients are given in Table 2.  

To simulate all presynaptic dynamics, not only steady state but also transient solutions for each concentration of Eq.(\ref{diffusion}) are necessary. To reduce its computational complexity, we use analytic solution of the equation;
\begin{eqnarray}
y_o(t,r)=y_0+\frac{\alpha \eta \Phi}{8(\pi d)^{\frac{3}{2}}} T \int^{t}_{0}\frac{ e^{-\frac{r^2}{4ud}}}{u\sqrt{u}} e^{\Lambda u} du  T^{-1}
\begin{pmatrix}
0\\ \cdot \\0\\1
\end{pmatrix}
  ~~~~~~~~~ ({\rm channel \ open}), \label{analytic1}\\
y_c(t,r)=y_0-\frac{\beta \eta \Phi}{8(\pi d)^{\frac{3}{2}}} T \int^{t}_{0}\frac{ e^{-\frac{r^2}{4ud}}}{u\sqrt{u}} e^{\Lambda u} du  T^{-1}
\begin{pmatrix}
0\\ \cdot \\0\\1
\end{pmatrix}
  ~~~~~~~~~({\rm channel \ close}),
\label{analytic2}
\end{eqnarray}

where $y_0$ is the concentration when a calcium channel is just open or closed, $y_{\infty}$ is the saturated concentration, $\alpha=\frac{y_{\infty}-y_0}{y_{\infty}}$, $\beta=\frac{y_{0}}{y_{\infty}}$, $\eta=0.19\sqrt{\frac{r}{r_0}-1}+1$, and $T$ is the matrix consisting of the eigenvectors of $A$. 
In the processing of obtaining the above equations, $D$ was replaced by $d_{Ca^{2+}}{\bf I}$, where $d_{Ca^{2+}}$ is the Calcium ion diffusion constant (Table 2) and ${\bf I}$ is the unit matrix. 
The use of the analytic solutions are guaranteed by our finding that there is no significant difference between the numerical solution from Eq.(\ref{diffusion})  and the analytic solution in Eqs.(\ref{analytic1},\ref{analytic2}).
The source parameter $\Phi$ is obtained by the experimental results\cite{Tay}, showing that $[Ca^{2+}]$ near the mouth of the channel follows a relationship of $700\ \mu M/pA$.

The solutions of Eq. (\ref{diffusion}) represent calcium flows when a calcium channel opens and close to ($y_o\ and\ y_c$) in the open boundary condition. 
Calculating the above integrals numerically takes a long time for long-time simulation if we perform the integration whenever necessary, so we use the following analytic formula;
$
\int^{t}_{0}\frac{\exp(-\frac{r^2}{4ud}+\lambda u)}{u\sqrt{u}}du = \Bigg[-\sqrt{\frac{\pi}{\tau}}(g_+(u)-g_-(u))\Bigg]^{t}_{0},
$
where $\tau=r^2/d$ and $g_{\pm}(u)=e^{\pm \sqrt{-\lambda \tau}}\left({\rm erf}\left(\frac{\sqrt{\tau}\pm2\sqrt{-\lambda}u}{2\sqrt{u}}\right)\mp1\right)$. 

The calcium channel clusters in auditory hair cells are found located under a ribbon structure and the height is around 100 nm \cite{Rutherford,Neef1}. The diffusion coefficient $d$ is actually the diagonal matrix, but there is no significant difference between the result from simulation and analytic solution, which switched $D$ as scalar value $d_{Ca^{2+}}$ (See Table 2). Comparing the simulation of Eq. (\ref{diffusion}) and the analytic solution, there is minor difference in the maximum value of the calcium concentration (See Supplementary Material)). For simulating Eq. (\ref{diffusion}), calcium channels modelled as shown in Fig. \ref{figure2} a), and the channels are spaced about by $20 nm$ distance. The total size of the simulation environment is $x:y:z=26:40:6$ with a spatial interval of $dx=20$ nm, which refers to the size of $500\times780\times100\ nm^3$. The bottom and ceiling are blocked with hair cell membrane and the ribbon structure. To reduce the difference in Fig.\ref{figure3}, we insert additional term $\eta$ to concern the ceiling.

\subsection{Sensor-ion interaction}

In the presynaptic zone, the vesicles are attached to the membrane, but not fused directly. There are protein complex, called SNARE, between the vesicle and the membrane, which causes vesicle fusion. Among them, synaptitagmin functions as a calcium sensor and tightly regulates SNARE zipping \cite{Lyshevski}. Therefore, the rate of vesicle fusion varies depending on the surrounding calcium concentration. However, it is very difficult to simulate all the movement of these proteins. In Ref. \cite{Beutner}, the calcium concentration inside the hair cell was controlled and fitted using a kinetic model. In this model, the calcium sensor binds to five calcium ions and subsequently causes vesicle fusion (Fig. \ref{figure2}c) ).

To achieve continuous vesicle fusion, a mechanism is needed to refill the vesicle after fusion. To get the replenishment rate, we used the sustained rate in Ref. \cite{Pangrsic}. The different point between previous model and ours is that vesicles are immediately fused when the sensor gets the 5th calcium ion. In the 5 calcium binding model, vesicles are fuses with a specific probability when the sensor combined with 5 calcium ions. And the parameter is fitted as EPSC is generated immediately in the fuse state. However, in our model, the process in which the vesicle releases neurotransmitters and EPSC is generated is also simulated separately, which causes problems in this part. So we assumed that the moment five calcium ions are attached, fusion occurs.

\begin{figure} [hbt!]
\centering
\includegraphics[width=1\textwidth]{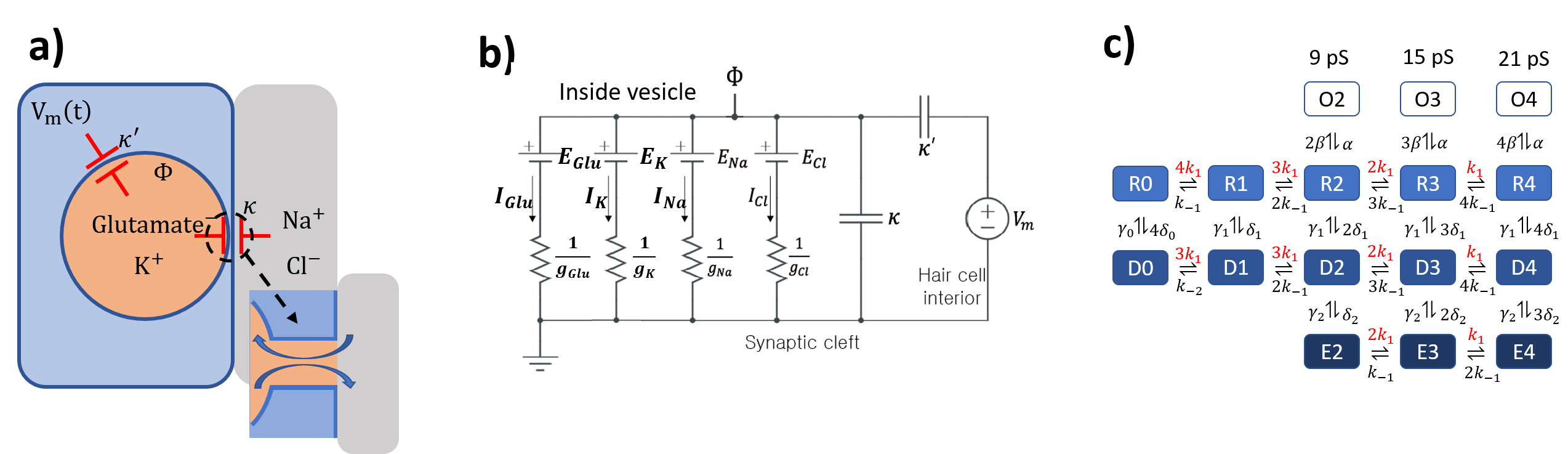}
\caption{
a) Schematic of vesicle fusion model. There are glutamate and potassium inside the vesicle, and sodium and chlorin in the synaptic cleft.
b) Circuit model of ion exchange.
c) 16-state kinetic model for AMPA receptor. The value of rate constants follow the Table 4.
}
\label{figure5}
\end{figure}

\subsection{Glutamate diffusion and its receptor}

The vesicle fusion process directly affects the subsequant glutamate release into the synaptic cleft (Fig. \ref{figure5}a ). To investigate the relationship between vesicle fusion kinetics and EPSC properties, we simulate the effects of pore opening rate and vesicle size on EPSC amplitude.

In our model, the expansion of the fusion pore is described by the following equation;
\begin{eqnarray}
r_{pore} = r_0(1-e^{-\gamma(t-t_0)}),
\label{pore_dyn}
\end{eqnarray}

where $\gamma$ denotes the pore opening rate, $r_{0}$ is the maximum pore radius, and $t_0$ is the onset time of vesicle fusion. As shown in Fig. \ref{figure4}a , the EPSC amplitude increases with increasing $\gamma$ up to approximately $2000 s^{-1}$, beyond which further increases in $\gamma$ result in minimal changes in amplitude. For an average vesicle radius of 16.5 nm, the typical EPSC amplitude reaches about 150 pA.

Furthermore, we analyze how the number of simultaneously fusing vesicle affects the EPSC amplitude (Fig. \ref{figure4}b ). The fusion of 3 vesicles can generate an EPSC amplitude of approximately 250 pA, corresponding well to the experimentally observed average amplitude\cite{Nikolai}. We also examined the relationship between vesicle fusion timing and EPSC amplitude (Fig. \ref{figure4}c ), and found that the delay time between vesicle fusions decreases slightly beyond $\gamma=2000s^{-1}$.

A model for discharge of charged excitatory neurotransmitter has been proposed for ACh-containing spherical vesicle, which contains $ACh^+, Na^+,$ and $Cl^-$ \cite{Khanin}. It is shown theoretically in this model that the ion interchange of the same sign of co-discharge of ions with the opposite sign are necessary for rapid vesicle discharge ($\approx 100\ {\mu s}$). For auditory haircell, negatively charged glutamates are relevant rather than acetycholine. In addition, our model for hair cell synapse has further structure due to the fact that membrane voltage is controlled by the influx of $K^+$ from hair bundle gated-channels.
We introduce an electronic circuit for the discharge of the charged neurotransmitters in a vesicle of hair cell synapse (Fig. \ref{figure5}b ). The dynamics ion concentration $C_j\ (j=Glu,Na,Cl,K)$ and the transpore potential $\Phi$ are governed by the Nernst equilibrium potential

\begin{eqnarray}
E_j = \frac{RT}{z_j F} ln\frac{C^{(e)}_j}{C_j},
\label{fusion_E}
\end{eqnarray}

where $z_j$ is the charge of species j with $z_{Glu}=z_{Cl}=-1$ and $z_{Na}=z_{K}=1$. $F$ is the Faraday constant, $R$ is the universal gas constant, and $T$ is the room temperature. The bias voltage for the pore is $(\Phi-E_j)$ where its corresponding conductance is $g_j$. The governing equation of motion are written as

\begin{eqnarray}
I_j = g_j(\Phi-E_j),
\label{fusion_I}
\end{eqnarray}

\begin{eqnarray}
\frac{dC_j}{dt}=-\frac{g_j}{z_j FV}(\Phi-E_j),
\label{fusion_dC}
\end{eqnarray}

\begin{eqnarray}
\frac{d\Phi}{dt} = -\frac{I_\Sigma}{\kappa+\kappa'}+\frac{\kappa'}{\kappa+\kappa'}\frac{dV_m(t)}{dt},
\label{fusion_Phi}
\end{eqnarray}

where $\kappa$ is the maximum pore capacitance, $\kappa'$ is a capacitance between a vesicle and haircell inside, and $I_\Sigma=I_{Glu}+I_{Na}+I_{Cl}+I_{K}$. $V$ is the vesicle volume, and $V_m$ is the membrane voltage. The parameters used for simulating vesicle fusion and glutamate release are summarized in Table 3.
The conductance for each ion can be modeled as Ref. \cite{Khanin}

\begin{eqnarray}
g_j=g_{pore}\frac{C_j}{C_j+C_{h}}=\frac{A_p}{\rho l}\biggl(\frac{C_j}{C_j+C_{h}}\biggr),
\label{fusion_g}
\end{eqnarray}

where $A_p$ is the pore size, $\rho$ is the pore resistivity, and $l$ is the pore length. Here the concentration dependence was introduced to describe the decreasing rate of ion's collision with pore and $C_h$ is given roughly the half of the initial concentration of species j \cite{Khanin}.
The passage of the particles through the pore is a highly complex, which will be influenced by dielectric force, local dipoles, and local charges that would make peaks and valleys in the potential profile \cite{Hille}. There has been no universal view on the pore. In synaptic vesicle fusion, the vesicle must be within a few nanometers of the target membrane to initiate the fusion process. Once the vesicle are in place, they wait for $Ca^{2+}$ to enter the cell membrane. $Ca^{2+}$ binds to one of the neuron's specific proteins (Synaptotagmin) that trigger complete fusion of the target membrane and vesicle \cite{Roux, Brady}.

When vesicles releases glutamate ions on the synaptic cleft, they diffuse to the postsynapse and bind with glutamate-AMPA receptors. Unlike the presynapse, there are no buffer in the synaptic cleft. Glutamate transporters remove glutamate and reuptake into the IHC, but there are on supporting cells near the IHC, which is far from the AMPA receptor cluster. So, we neglect the effect of the glutamate transporter.

Glutamate is released from the presynaptic terminal in the synaptic cleft. AMPA receptors are the major glutamate receptors in the inner hair cell. The receptors mediate fast excitatory transmission and rapid desensitization of these receptors can shape the decay of synaptic currents and limit the fidelity of high-frequency synaptic transmission. For the AMPAR model, there are 4 hypothetical states of open. To mimic rapid desensitization of receptors, the model is combined with desensitization states. Finally, the dynamics of AMPA receptors are governed by the kinetic schema for the 16-state AMPA receptor model, depicted in binding of Glutamate to the receptor (Fig. \ref{figure5}c ). We use a 16 state AMPA receptor model with desensitization state (D,E) and 3 open states (O).

\section{Conclusions}
In this study, we simulated the full dynamics of auditory synaptic fusion based on parameters derived from a wide range of experimental findings. The simulation encompasses the gating of calcium channels, the subsequent diffusion of calcium ions, and the interaction between calcium ions and calcium sensors, all modeled with high fidelity. Additionally, we simulated the glutamate release process in detail, which enabled us to replicate the experimentally observed excitatory postsynaptic currents (EPSCs).

There has been ongoing debate regarding the characteristics of EPSCs observed in auditory synapses. One central question concerns the origin of unusually large EPSCs: whether they result from the simultaneous fusion of multiple vesicles (multivesicular release) or from the fusion of a single, large vesicle (univesicular release). Another point of contention lies in the shape of the EPSCs. Both monophasic EPSCs, which feature a single, prominent peak, and multiphasic EPSCs, which display irregular, fluctuating profiles, are observed experimentally. The mechanisms underlying these distinct EPSC waveforms remain unclear.

Our findings suggest that it is physically implausible to account for multiphasic EPSCs through univesicular release mechanisms alone. While multivesicular release more readily explains the occurrence of multiphasic EPSCs, it struggles to account for the large amplitudes observed in monophasic EPSCs. In our study, we considered the possibility that calcium channels may experience attractive forces that promote their clustering. Our simulations demonstrate that strong calcium channel clustering can enable multivesicular release to produce large-amplitude, monophasic EPSCs.

Therefore, the phenomena observed in EPSCs at the auditory synapse can be more naturally explained by the multivesicular release mechanism, particularly when calcium channel clustering is taken into account.

\begin{acknowledgments}
We thank Prof. Eungyoung Yi, Prof. Sunghwa Hong, and Prof. Il Joon Moon for helpful discussion. This research was supported by Basic Science Research
Program through the National Research Foundation of
Korea(NRF) funded by the Ministry of Education(RS-
2023-00246572). 
\end{acknowledgments}


\newpage

\begin{table}
\begin{center}
\begin{tabular}{cccc}
\hline\hline
Parameter & Value & Unit \\
\hline

$g_{max}$ & 3.8 & nS &\\
\hline

$V_{rev}$ & -23.4 & mV \\
\hline

$V_{\frac{1}{2}}$ & 38 & mV \\
\hline

$S$ & 8 & mV \\

\hline\hline
\end{tabular}
\caption{Parameters for calcium channel gating \cite{Marcotti}.}\label{table1}
\end{center}
\end{table}

\newpage
\begin{table}[p]
  \centering
\begin{tabular}{ p{3.7cm}l }
\hline
  Parameter & Value \\
 \hline
  $[CR2T]_r, [CR2R]_r$ & 0.036\ mM \\
  $[CR1]_r$ & 0.018\ mM \\
  $[CB]_r$ & 0.232\ mM \\
  $[PV]_r$ & 0.188\ mM \\
  $[ATP]_r$ & 0.165\ mM \\
  $[Ca^{2+}]_r$ & $5\times10^5$\ mM \\
  $d_{Ca^{2+}}$ & $2.2\times10^{-10}$\ $s/m^2$ \\
  \hline
 \end{tabular}
 \caption{List of the parameter values which were used in calcium diffusion. These are taken from Ref(\cite{Wong}).}\label{table2}
\end{table}

\newpage
\begin{table}[p]
  \centering
\begin{tabular}{ p{2cm} p{6cm} p{3cm} l }
\hline
  Parameter & Definition & Value & References \\
 \hline
  $C_{\rm Glu}$ & Initial glutamate concentration & 10$^{-3}$ mM (Outside) \ 180 mM (Inside) &\cite{Scimemi}  \\
  $C_{\rm Na}$ & Initial sodium concentration & 110 mM (Outside) \ 1 mM (Inside)  & \cite{Zylbertal} \\
  $C_{\rm Cl}$ & Initial chloride concentration   & 110 mM (Outside) \ 1 mM (Inside)  & \cite{Schulte} \\
  $C_{\rm K}$ & Initial potassium concentration   & 175 mM (Outside) \ 2.8 mM (Inside)  & \cite{Schulte} \\
  $r_V$ & Vesicle diameter & 40  $\text{nm}$ & \cite{Khimich} \\
  $\kappa$ & Maximum pore capacitance & 42.4 aF & \cite{Khanin} \\
  $\kappa'$ & Capacitance between a vesicle and haircell inside & $1.5\kappa$ & \cite{Wu} \\
  $g(A_{m})$ & Maximum conductance & 250 pS & \cite{Khanin} \\
  $\Phi(0)$ & Initial potential & -260 mV & \cite{Khanin} \\
  $T$ & Room temperature & 298 K & \\
  \hline
 \end{tabular}
 \caption{List of the parameter values for vesicle fusion simulation}\label{table3}
\end{table}

\newpage
\begin{table}[p]
  \centering
\begin{tabular}{ p{5cm}l }
\hline
  Parameter & Value\\
 \hline
  $k_1, k_2$ & 10\ $mM^{-1}$\ $ms^{-1}$ \\
  $k_{-1}$ & 7\ $ms^{-1}$ \\
  $k_{-2}$ & 0.00041\ $ms^{-1}$ \\
  $\delta_0$ & 3.3$\times 10^{-6}$\ $ms^{-1}$ \\
  $\delta_1$ & 0.42\ $ms^{-1}$ \\
  $\delta_2$ & 0.2\ $ms^{-1}$ \\
  $\gamma_0$ & 0.001\ $ms^{-1}$ \\
  $\gamma_1$ & 0.017\ $ms^{-1}$ \\
  $\gamma_2$ & 0.035\ $ms^{-1}$ \\
  $\alpha$ & 0.3\ $ms^{-1}$ \\
  $\beta$ & 0.55\  $ms^{-1}$ \\
  \hline
 \end{tabular}
 \caption{List of the parameter values for AMPA receptors \cite{Bouteiller, Robert}.}\label{table4}
\end{table}

\end{document}